\documentclass[aps,pre,twocolumn,groupedaddress,10pt,showpacs,superscriptaddress]{revtex4-1}
\usepackage{amsmath,amssymb}
\usepackage{graphicx,color}
\usepackage{rotating,array,tabularx,booktabs}
\newcolumntype{Y}{>{\centering\arraybackslash}X}

\newcommand{\dif}{\mathrm{d}}%
\newcommand{\Diag}{\operatorname{diag}}%

\newcommand{\ii}{i}%
\newcommand{\Nabla}{\vec{\nabla}}%
\newcommand{\R}{\mathbb{R}}%
\newcommand{\abs}[1]{\lvert#1\rvert}%
\newcommand{\ie}{i.\,e.}%

\begin{document}
\title{The self-propelled Brownian spinning top: dynamics of a biaxial swimmer at low Reynolds numbers}

\author{Raphael Wittkowski}\author{Hartmut L{\"o}wen}
\affiliation{Institut f{\"u}r Theoretische Physik II, Weiche Materie,
Heinrich-Heine-Universit{\"a}t D{\"u}sseldorf, D-40225 D{\"u}sseldorf, Germany}
\date{\today}

\begin{abstract}
Recently, the Brownian dynamics of self-propelled (active) rod-like particles was explored to model the motion of
colloidal microswimmers, catalytically-driven nanorods, and bacteria.
Here, we generalize this description to biaxial particles with arbitrary shape and derive the corresponding Langevin equation 
for a self-propelled Brownian spinning top.
The biaxial swimmer is exposed to a hydrodynamic Stokes friction force at low Reynolds numbers, 
to fluctuating random forces and torques as well as to an external and an internal (effective) force and torque. 
The latter quantities control its self-propulsion.
Due to biaxiality and hydrodynamic translational-rotational coupling, the Langevin equation 
can only be solved numerically. In the special case of an orthotropic particle in the absence of external forces and torques, the 
noise-free (zero-temperature) trajectory is analytically found to be a circular helix. 
This trajectory is confirmed numerically to be more complex in the general case involving a transient irregular motion
before ending up in a simple periodic motion. By contrast, if the external force vanishes, no transient regime
is found and the particle moves on a \emph{superhelical} trajectory. For orthotropic particles,
the noise-averaged trajectory is a generalized \emph{concho-spiral}. We furthermore study the reduction
of the model to two spatial dimensions and classify the noise-free trajectories completely finding
circles, straight lines with and without transients, as well as cycloids and arbitrary periodic trajectories.
\end{abstract}

% Colloids: 82.70.Dd
% Brownian motion: 05.40.Jc

\maketitle

%**************************************************************************
%**************************************************************************

\section{\label{sec:introduction}Introduction}
In the traditional description of colloidal particles, their shape is assumed to be either spherical or rod-like
\cite{Dhont1996,DoiE2007,Loewen2001}, or -- in other words -- the particles are 
either isotropic or uniaxial, \ie, rotationally symmetric around a figure axis 
\cite{FradenMCM1989,GrafL1999b,JohnsonvKvB2005,RoordavDPGvBK2004,WierengaLP1998}. 
In the isotropic case, the location of the particle is described by its center-of-mass position, 
while for uniaxial particles an additional unit vector is needed to describe its orientation. 
Using various preparation techniques, by now, it is possible to prepare colloidal particles with a more complex shape 
than spherical and uniaxial in a controlled way
\cite{GlotzerS2007,ManoharanEP2003,SolomonZOSDSBGM2010,QuillietZRvBI2008,KraftVvKvBIK2009,Lubensky2000}. 
In particular, it is possible to synthesize particles with an orthotropic shape as, for example, board-like colloids
\cite{vandenPolPTWBV2009}. However, while the theory of Brownian dynamics for spherical and uniaxial colloidal particles is quite advanced
\cite{Dhont1996,DoiE2007,ImperioRZ2008,Loewen1994b}, much less is known for biaxial particles which need an additional angle to describe 
their location in space (on top of the center-of-mass position and the orientational unit vector). 
This additional degree of freedom complicates the description of Brownian motion considerably \cite{CaldererFW2004,WhiteCH2001}.

A second recent development in colloid science is to make the colloidal particles self-propelling such that 
they are moving in space on their own (so-called \emph{active} particles) \cite{TonerTR2005,GaugerS2006}. 
There are numerous examples of self-propelling colloids including catalytically driven nanorods 
\cite{DharFWMPS2006,WaltherM2008,ErbeZBKL2008,PopescuDO2009} and
thermo-gradient driven Janus-particles \cite{VolpeBVKB2011}, not to speak about other realizations of swimmers
in vibrating granulates \cite{vanTeeffelenZL2009}, magnetic beads \cite{DreyfusBRFSB2005}, and real biological systems 
\cite{BergT1990,DiluzioTMGWBW2005,LaugaDWS2006,HillKMK2007,ShenoyTPT2007,SchmidtvdGBWHF2008,RiedelKH2005,Woolley2003,FriedrichJ2008},
where, for example, protozoa use cilia, flagella, helical traveling waves, and protoplasmic flow for locomotion \cite{JahnV1972}. 
The simplest description of self-propulsion is modelled by an internal effective force which provides an 
constant propulsion mechanism on top of the Brownian motion of the particle 
\footnote{This should not be confused with the basic fact that a swimmer is force-free and torque-free.}. 
In two spatial dimensions, it was shown recently that the simultaneous action of an internal force and an internal torque leads to circle-swimming
\cite{vanTeeffelenL2008,vanTeeffelenZL2009} and in three spatial dimensions, the behavior of the mean square displacement 
was calculated analytically \cite{tenHagenvTL2009,tenHagenvTL2011,HowseJRGVG2007}. 

All considerations for Brownian swimmers were done hitherto for uniaxial particles. 
We are not aware of any study for biaxial swimmers except for a recent modeling by Vogel and Stark \cite{Vogel_Stark}.
This is important for at least two reasons: first, real particles are in general not uniaxial and therefore
the effect of biaxiality needs to be studied. Second, it is of general importance to generalize the equations
of a "Brownian spinning top" towards self-propulsion in order to understand and predict its motion on a fundamental level. 
Surprisingly, while the spinning top in an external field governed by the Newtonian (or Eulerian) equations is a standard reference 
model in classical mechanics, there are only few studies for overdamped Brownian motion of a passive spinning top \cite{FernandesdlT2002,MakinoD2004}.

In this paper, we derive the Langevin equation for a biaxial self-propelled particle with arbitrary shape
that may even be screw-like, implying a translational-rotational coupling \cite{Brenner1965,Brenner1967}.
The biaxial swimmer is exposed to a hydrodynamic Stokes friction force at low Reynolds numbers, 
to fluctuating random forces and torques, as well as to an external and an internal (effective) force and torque. 
The internal forces and torques control the translational and angular propulsion velocity and are constant in the body-fixed frame
while the external force and torque is constant in the (space-fixed) lab frame.

Due to biaxiality and hydrodynamic translational-rotational coupling, the Langevin equation 
can only be solved numerically. In the special case of an orthotropic particle, 
which has no translational-rotational coupling, the noise-free (zero-temperature) trajectory is analytically found to be a 
circular helix in the absence of any external force and torque. 
This trajectory is confirmed numerically to be much more complex in the general case. Typically, there is 
an irregular transient motion before the particle ends up in a simple periodic motion. 
Such helix-like trajectories are indeed typical for swimming microorganisms \cite{JahnV1972,SchulzJ2001}.
By contrast, if the external force vanishes, no transient motion
shows up and the particle exhibits  a \emph{superhelical} trajectory. For orthotropic particles,
the noise-averaged trajectory is studied numerically and found to be a generalized \emph{concho-spiral}
\cite{Boyadzhiev1999} with a "snail-shell" structure.
We furthermore study the reduction of the model to two spatial dimensions where there is only one orientational angle 
and the internal and external torques can be combined to a single effective torque. 
In this simpler two-dimensional limit, we classify the noise-free trajectories completely.
Circles are found in the absence of the external force. 
In general, the trajectories are straight lines with and without transients as well as cycloids and arbitrary periodic curves.

This paper is organized as follows: in Sec.\ \ref{sec:derivation} we present the Langevin equation for a general self-propelled
rigid biaxial Brownian particle in an unbounded viscous fluid with low Reynolds number being at rest at infinity. 
Section \ref{sec:SAS} is dedicated to special analytical solutions and Sec.\ \ref{sec:NC} to more general numerical calculations 
for the Langevin equation. Finally, we conclude and give an outlook in Sec.\ \ref{sec:conclusions}.

\section{\label{sec:derivation}Langevin equation}
In this section, we describe the Brownian motion of a biaxial self-propelled particle suspended in an unbounded viscous fluid at rest at infinity
for low Reynolds numbers. 
It is assumed that the colloidal particle is rigid and has a constant mass density.
The motion of this particle is characterized by the translational center-of-mass velocity $\dot{\vec{r}}=\dif\vec{r}/\dif t$ 
with the center-of-mass position $\vec{r}(t)$ and the time variable $t$ as well as by the instantaneous angular velocity $\vec{\omega}(t)$. 
The Brownian motion of colloidal particles with arbitrary shape involves a coupling between the translational and the rotational 
degrees of freedom, which was described theoretically, for example, by Brenner \cite{Brenner1965,Brenner1967}. 
In 2002, Fernandes and de la Torre \cite{FernandesdlT2002} have proposed a corresponding Brownian dynamics simulation
algorithm for the motion of a passive rigid particle of arbitrary shape. The underlying equations of motion were generalized to an imposed external flow
field for the surrounding fluid by Makino and Doi \cite{MakinoD2004}.

Here we appropriately generalize this description to \textit{internal} degrees of freedom, \ie, to a self-propelled
biaxial particle that experiences an internal effective force $\vec{F}_{0}$ and torque $\vec{T}_{0}$, which are
both constant in the body-fixed system. This models a biaxial microswimmer. 
Of course, a swimmer is in principle force-free and torque-free but the internal forces and 
torques are meant to be effective quantities which govern the propulsion mechanism of the particle.
Using a compact notation, we introduce the basic completely overdamped 
\textit{Langevin equation for three spatial dimensions} 
\begin{equation}
\begin{split}
\vec{\mathfrak{v}}&=\beta\mathcal{D}(\vec{\mathfrak{x}})\big(\mathcal{R}^{-1}(\vec{\mathfrak{x}})\vec{K}_{0}
-\Nabla_{\vec{\mathfrak{x}}}U(\vec{\mathfrak{x}})+\mathcal{R}^{-1}(\vec{\mathfrak{x}})\vec{k}\big)\\
&\quad+\Nabla_{\vec{\mathfrak{x}}}\cdot\mathcal{D}(\vec{\mathfrak{x}})
\end{split}
\label{eq:LangevinGLG}
\end{equation}
for a self-propelled Brownian spinning top.
Let us first explain the notation step by step. 
The biaxial particle has the position $\vec{r}=(x_{1},x_{2},x_{3})$ and the orientation $\vec{\varpi}=(\phi,\theta,\chi)$, 
which is given in Eulerian angles
\footnote{As there is no uniqueness in the definitions of the Eulerian angles $\vec{\varpi}=(\phi,\theta,\chi)$, 
we use for convenience the popular convention of Gray and Gubbins \cite{GrayG1984}, which is equivalent to the second convention 
of Schutte \cite{Schutte1976}. 
This convention has the advantage, that it is a direct generalization of the spherical coordinates $(\theta,\phi)$ 
that are identical with the first two Eulerian angles $\phi$ and $\theta$, while the third angle $\chi$ describes the rotation around the 
axis that is defined by $\theta$ and $\phi$ in the spherical coordinate system.}.
We summarize translational and rotational degrees of freedom by a compact $6$-dimensional vector
$\vec{\mathfrak{x}}=(\vec{r},\vec{\varpi})$, which obviously involves a generalized velocity  
$\vec{\mathfrak{v}}=(\dot{\vec{r}},\vec{\omega})$ (with $\vec{\omega}$ denoting the angular velocity) 
and a gradient $\Nabla_{\vec{\mathfrak{x}}}=(\Nabla_{\vec{r}},\Nabla_{\vec{\varpi}})$. The latter gradient is composed both of
 the usual translational gradient operator $\Nabla_{\vec{r}}=(\partial_{x_{1}},\partial_{x_{2}},\partial_{x_{3}})$ 
acting on the Cartesian coordinates of $\vec{r}$ and the rotational gradient operator 
$\Nabla_{\vec{\varpi}}=\ii\hat{\mathrm{L}}$ \cite{GrayG1984},
given by the product of the imaginary unit $\ii$ and the angular momentum operator 
$\hat{\mathrm{L}}=(\mathrm{L}_{1},\mathrm{L}_{2},\mathrm{L}_{3})$
in Eulerian angles.
In the space-fixed coordinate system, the angular momentum operator $\hat{\mathrm{L}}$ is given in Eulerian angles by \cite{GrayG1984}
{\allowdisplaybreaks\begin{align}%
\begin{split}%
\ii\:\! \mathrm{L}_{1}&=-\cos(\phi)\cot(\theta)\:\!\partial_{\phi}-\sin(\phi)\:\!\partial_{\theta}\\
&\quad\,+\cos(\phi)\csc(\theta)\:\!\partial_{\chi}\;,
\end{split}\\
\begin{split}%
\ii\:\! \mathrm{L}_{2}&=-\sin(\phi)\cot(\theta)\:\!\partial_{\phi}+\cos(\phi)\:\!\partial_{\theta}\\
&\quad\,+\sin(\phi)\csc(\theta)\:\!\partial_{\chi}\;,
\end{split}\\
\begin{split}%
\ii\:\! \mathrm{L}_{3}&=\partial_{\phi}\;.
\end{split}%
\end{align}}%
The angular velocity $\vec{\omega}$ is expressed in Eulerian angles by
\begin{equation}
\vec{\omega}=\mathrm{M}(\vec{\varpi})\:\!\dot{\vec{\varpi}}
\end{equation}
with the tensor \cite{Schutte1976}
{\allowdisplaybreaks\begin{align}%
\mathrm{M}(\vec{\varpi})&=
\begin{pmatrix}
0 & -\sin(\phi) & \cos(\phi)\sin(\theta) \\
0 & \cos(\phi)  & \sin(\phi)\sin(\theta) \\
1 & 0           & \cos(\theta)
\end{pmatrix} , \\
\mathrm{M}^{-1}(\vec{\varpi})&=
\begin{pmatrix}
-\cos(\phi)\cot(\theta) & -\sin(\phi)\cot(\theta) & 1 \\
-\sin(\phi)             & \cos(\phi)              & 0 \\
\cos(\phi)\csc(\theta)  & \sin(\phi)\csc(\theta)  & 0
\end{pmatrix}%
\end{align}}%
and the Eulerian angular velocities $\dot{\vec{\varpi}}=\dif\vec{\varpi}/\dif t$.
Furthermore, in Eq.\ \eqref{eq:LangevinGLG} the compact notation $\vec{K}_{0}=(\vec{F}_{0},\vec{T}_{0})$ for the generalized force
is used, which combines the effective propulsion force and torque.
In general, the biaxial particles is also exposed to an external potential $U(\vec{\mathfrak{x}})$
giving rise to an external force $\vec{F}_{\mathrm{ext}}=-\Nabla_{\vec{r}}U$ and an external torque $\vec{T}_{\mathrm{ext}}=-\Nabla_{\vec{\varpi}}U$,
which we both consider to be constant in the sequel in order to keep the model simple. 

The $6\!\times\!6$-dimensional matrix $\mathcal{R}^{-1}(\vec{\mathfrak{x}})$ is associated with the geometric transformation 
from the body-fixed frame to the lab frame. Its inverse $\mathcal{R}(\vec{\mathfrak{x}})$ is given as the block diagonal matrix 
\begin{equation}
\mathcal{R}(\vec{\mathfrak{x}})=\Diag\big(\mathrm{R}(\vec{\varpi}),\mathrm{R}(\vec{\varpi})\big)
\end{equation}
with the submatrices 
\begin{equation}
\begin{split}
\mathrm{R}(\vec{\varpi})&=\mathrm{R}_{3}(\chi)\,\mathrm{R}_{2}(\theta)\,\mathrm{R}_{3}(\phi)\;,\\
\mathrm{R}^{-1}(\vec{\varpi})&=\mathrm{R}^{\mathrm{T}}(\vec{\varpi})=\mathrm{R}_{3}(-\phi)\,\mathrm{R}_{2}(-\theta)\,\mathrm{R}_{3}(-\chi)\;,
\end{split}
\end{equation}
where the elementary rotation matrices $\mathrm{R}_{i}(\varphi)$ describe a 
clockwise rotation (when looking down the axes) 
around the $i$th Cartesian axis by the angle $\varphi$ for $i\in\{1,2,3\}$:
{\allowdisplaybreaks\begin{align}%
\mathrm{R}_{1}(\varphi)&=
\begin{pmatrix}
1              & 0              & 0              \\
0              & \cos(\varphi)  & \sin(\varphi)  \\
0              & -\sin(\varphi)  & \cos(\varphi)
\end{pmatrix} , \\
\mathrm{R}_{2}(\varphi)&=
\begin{pmatrix}
\cos(\varphi)  & 0              & -\sin(\varphi) \\
0              & 1              & 0              \\
\sin(\varphi)  & 0              & \cos(\varphi) 
\end{pmatrix} , \\
\mathrm{R}_{3}(\varphi)&=
\begin{pmatrix}
\cos(\varphi)  & \sin(\varphi)  & 0              \\
-\sin(\varphi) & \cos(\varphi)  & 0              \\
0              & 0              & 1
\end{pmatrix} .
\end{align}}%
The particle shape and its hydrodynamics enter in the generalized short-time diffusion (or inverse friction)
tensor $\mathcal{D}(\vec{\mathfrak{x}})$. This can be expressed as the $6\!\times\!6$-dimensional matrix
\begin{equation}
\begin{split}
\mathcal{D}(\vec{\mathfrak{x}})&=
\begin{pmatrix}
\mathrm{D}^{\mathrm{TT}}(\vec{\varpi}) & \mathrm{D}^{\mathrm{TR}}(\vec{\varpi}) \\
\mathrm{D}^{\mathrm{RT}}(\vec{\varpi}) & \mathrm{D}^{\mathrm{RR}}(\vec{\varpi}) 
\end{pmatrix}\\
&=\frac{1}{\beta\eta}\,\mathcal{R}^{-1}(\vec{\mathfrak{x}})\,\mathcal{H}^{-1}\,\mathcal{R}(\vec{\mathfrak{x}}) \;,
\end{split}
\end{equation}
where $\mathrm{D}^{\mathrm{TT}}(\vec{\varpi})$, $\mathrm{D}^{\mathrm{TR}}(\vec{\varpi})=(\mathrm{D}^{\mathrm{RT}}(\vec{\varpi}))^{\mathrm{T}}$, 
and $\mathrm{D}^{\mathrm{RR}}(\vec{\varpi})$ are $3\!\times\!3$-dimensional submatrices that correspond to pure translation, 
translational-rotational coupling, and pure rotation, respectively
\footnote{For experimental investigations of the three-dimensional translational and rotational diffusion of biaxial colloidal particles, see, 
for example, Refs.\ \cite{HoffmannWHW2009,HunterEEW2011} and references therein.}.
Here, $\eta$ is the viscosity of the embedding fluid and $\beta=1/(k_{\mathrm{B}}T)$ 
with the Boltzmann constant $k_{\mathrm{B}}$ and the absolute (effective) temperature $T$ denotes the inverse thermal
energy.
The matrix $\mathcal{H}$ is constant. It only depends on the shape and the size of the Brownian particle and is composed of the symmetric translation 
tensor $\mathrm{K}$, the not necessarily symmetric coupling tensor $\mathrm{C}_{\mathrm{S}}$ 
\footnote{The coupling tensor becomes symmetric if one chooses the \emph{center of hydrodynamic reaction} as 
reference point $\mathrm{S}$ \cite{HappelB1991}.}, and the symmetric rotation tensor $\Omega_{\mathrm{S}}$ \cite{Brenner1967,HappelB1991}: 
\begin{equation}
\mathcal{H}=
\begin{pmatrix}
\mathrm{K} & \mathrm{C}^{\mathrm{T}}_{\mathrm{S}} \\
\mathrm{C}_{\mathrm{S}} & \Omega_{\mathrm{S}} \\
\end{pmatrix} .
\end{equation}
The Langevin equation \eqref{eq:LangevinGLG} thus involves altogether $21$ shape-dependent parameters.

Finally, $\vec{k}(t)=(\vec{f}_{0},\vec{\tau}_{0})$ denotes the stochastic force $\vec{f}_{0}(t)$ and torque $\vec{\tau}_{0}(t)$ 
due to thermal fluctuations, that act on the Brownian particle in body-fixed coordinates for $T>0$. 
This thermal noise $\vec{k}(t)$ is assumed to be Gaussian white noise with mean
\begin{equation}
\langle\vec{k}(t)\rangle=\vec{0}
\end{equation}
and correlation 
\begin{equation}
\langle\vec{k}(t_{1})\otimes\vec{k}(t_{2})\rangle=\mathcal{H}\frac{2\eta}{\beta}\,\delta(t_{1}-t_{2}) \;,
\end{equation}
where $\langle\,\cdot\,\rangle$ denotes the noise average.

We remark that the dynamics that is described by the Langevin equation depends on the definition of the stochastic contribution. 
In the present form, the Langevin equation is only valid, if the multiplicative noise is defined in the It\={o} sense \cite{Risken1996}.
Then, the additive drift term $\Nabla_{\vec{\mathfrak{x}}}\cdot\mathcal{D}(\vec{\mathfrak{x}})$ 
at the end of the Langevin equation guarantees that the solutions of the Langevin equation respect the Boltzmann distribution, 
when the system is in equilibrium at temperature $T$.
Other definitions than that of It\={o} like, for example, the Stratonovich formulation, require the addition of further special drift terms to 
the right-hand-side of the Langevin equation \cite{LauL2007}.
This circumstance is always relevant in the case of multiplicative noise, 
but the necessity of the adaptation of the Langevin equation to the 
definition of the stochastic noise has been missed in previous work \cite{FernandesdlT2002,MakinoD2004}.

\section{\label{sec:SAS}Special analytical solutions of the Langevin equation}
The Langevin equation \eqref{eq:LangevinGLG} represents a system of six coupled nonlinear stochastic 
differential equations \cite{Hassler2007} that cannot be solved analytically in general. 
There exist only a few analytical solutions for rather special situations.  
Several simple Langevin equations for self-propelled spherical or uniaxial particles in two or three spatial dimensions are known from literature
\cite{vanTeeffelenL2008,tenHagenvTL2009,tenHagenvTL2011} and appear to be special cases of our general Langevin equation \eqref{eq:LangevinGLG}. 
These Langevin equations are not as general and complicated as Eq.\ \eqref{eq:LangevinGLG} and can be solved analytically. 
Other analytically solvable special cases of Eq.\ \eqref{eq:LangevinGLG} are obtained for orthotropic particles in the absence of 
thermal fluctuations. 

\subsection{\label{subsec:LangevinIIID}Three spatial dimensions}
In the general three-dimensional case, it is not even possible to solve the Langevin equation \eqref{eq:LangevinGLG} analytically 
if the stochastic noise is neglected. Further simplifications that reduce the number of the degrees of freedom or diagonalize the
matrix $\mathcal{H}$ are necessary in order to obtain soluble cases.

\subsection{\label{subsec:LangevinIID}Two spatial dimensions}
\begin{table*}[ht]%
\renewcommand{\arraystretch}{1.5}%
\newcommand{\LZelleA}[1]{\parbox[c][0.16\linewidth][c]{\linewidth}{\raggedright#1}}%
\newcommand{\CZelleA}[1]{\parbox[c][0.16\linewidth][c]{\linewidth}{\centering#1}}%
\newcommand{\LZelleB}[1]{\parbox[c][0.32\linewidth][c]{\linewidth}{\raggedright#1}}%
\newcommand{\CZelleB}[1]{\parbox[c][0.32\linewidth][c]{\linewidth}{\centering#1}}%
\newcommand{\LZelleC}[1]{\parbox[c][0.48\linewidth][c]{\linewidth}{\raggedright#1}}%
\newcommand{\CZelleC}[1]{\parbox[c][0.48\linewidth][c]{\linewidth}{\centering#1}}%
\centering
\caption{\label{tab:Koeffizienten}Connection between the symmetry of the particle shape and the parameters \eqref{eq:Bp}-\eqref{eq:DR} 
in the Langevin equations \eqref{eq:LangevinGLGII} for two spatial dimensions.}
\begin{tabularx}{\linewidth}{|p{24mm}|Y|YY|Y|Y|}
\hline%\hline
\LZelleA{type of shape:} & \CZelleA{uniaxial} & \CZelleA{uniaxial} & \CZelleA{uniaxial} & \CZelleA{uniaxial} & \CZelleA{isotropic} \\  
\hline
\LZelleA{symmetries:} & \CZelleA{no symmetry} & \CZelleA{inflection symmetry} & \CZelleA{inflection symmetry} 
& \CZelleA{inflection symmetry} & \CZelleA{rotational symmetry} \\
\hline
\LZelleB{invariance properties:} & \CZelleB{---} & \CZelleB{$x_{1}\to-x_{1}$} & \CZelleB{$x_{2}\to-x_{2}$} 
& \CZelleB{$x_{1}\to-x_{1}$,\\$x_{2}\to-x_{2}$} & \CZelleB{$\phi\to\phi+\Delta\phi$ $\forall\;\Delta\phi\in[0,2\pi)$} \\
\hline
\LZelleC{shape-dependent\\parameters:\\[3pt]$D_{1}\neq 0$, $D_{\mathrm{R}}\neq 0$}
& \CZelleC{$B^{\parallel}_{\mathrm{I}}\neq 0$, $B^{\perp}_{\mathrm{I}}\neq 0$,\\$D_{2}\neq 0$, $D_{3}\neq 0$,\\
$D^{\parallel}_{\mathrm{C}}\neq 0$, $D^{\perp}_{\mathrm{C}}\neq 0$}
& \CZelleC{$B^{\parallel}_{\mathrm{I}}=0$, $B^{\perp}_{\mathrm{I}}\neq 0$,\\$D_{2}=0$, $D_{3}\neq 0$,\\ 
$D^{\parallel}_{\mathrm{C}}\neq 0$, $D^{\perp}_{\mathrm{C}}=0$}
& \CZelleC{$B^{\parallel}_{\mathrm{I}}\neq 0$, $B^{\perp}_{\mathrm{I}}=0$,\\$D_{2}=0$, $D_{3}\neq 0$,\\
$D^{\parallel}_{\mathrm{C}}=0$, $D^{\perp}_{\mathrm{C}}\neq 0$}
& \CZelleC{$B^{\parallel}_{\mathrm{I}}=0$, $B^{\perp}_{\mathrm{I}}=0$,\\$D_{2}=0$, $D_{3}\neq 0$,\\
$D^{\parallel}_{\mathrm{C}}=0$, $D^{\perp}_{\mathrm{C}}=0$}
& \CZelleC{$B^{\parallel}_{\mathrm{I}}=0$, $B^{\perp}_{\mathrm{I}}=0$,\\$D_{2}=0$, $D_{3}=D_{1}$,\\
$D^{\parallel}_{\mathrm{C}}=0$, $D^{\perp}_{\mathrm{C}}=0$} \\
\hline%\hline
\end{tabularx}%
\end{table*}%
An obvious simplification of the general Langevin equation \eqref{eq:LangevinGLG} is the restriction to two spatial dimensions.
A two-dimensional analog of the general Langevin equation can be obtained by choosing $x_{3}=0$, $\theta=\pi/2$, and $\chi=0$,
leaving only a single azimuthal angle $\phi$.
In analogy to the notation in Ref.\ \cite{vanTeeffelenL2008}, we further define 
$\mathcal{R}^{-1}(\vec{\mathfrak{x}})\vec{K}_{0}=(\vec{F}_{\mathrm{A}},0,0,0,M)$,
$\vec{k}(t)=(0,f_{\perp},f_{\parallel},-\tau,0,0)$, and 
$\mathcal{R}^{-1}(\vec{\mathfrak{x}})\vec{k}=(\vec{f},0,0,0,\tau)$, where 
$\vec{F}_{\mathrm{A}}=F_{\parallel}\hat{u}_{\parallel}+F_{\perp}\hat{u}_{\perp}$ is the internal driving force and $M$ is the internal 
driving torque. The orientation vector $\hat{u}_{\parallel}=(\cos(\phi),\sin(\phi))$ denotes the figure axis of the particle and
$\hat{u}_{\perp}=(-\sin(\phi),\cos(\phi))$ is its orthogonal complement.
Similarly, the vector $\vec{f}(t)$ denotes the stochastic force and $\tau(t)$ is the stochastic torque acting on the particle.
The stochastic force vector $\vec{f}(t)$ is decomposed like the internal driving force: 
$\vec{f}(t)=f_{\parallel}\hat{u}_{\parallel}+f_{\perp}\hat{u}_{\perp}$.
Moreover, the new two-dimensional position vector is $\vec{r}=(x_{1},x_{2})$, the corresponding gradient is 
$\Nabla_{\vec{r}}=(\partial_{x_{1}},\partial_{x_{2}})$, and the stochastic noise is characterized by 
$\tilde{\vec{\xi}}(t)=(f_{\perp},f_{\parallel},-\tau)$ with mean 
\begin{equation}
\langle\tilde{\vec{\xi}}(t)\rangle=\vec{0}
\end{equation}
and correlation 
\begin{equation}
\langle\tilde{\vec{\xi}}(t_{1})\otimes\tilde{\vec{\xi}}(t_{2})\rangle=\widetilde{\mathcal{H}}\:\!\frac{2\eta}{\beta}\,\delta(t_{1}-t_{2})\;,
\end{equation}
where $\widetilde{\mathcal{H}}=(\mathcal{H}_{ij})_{i,j=2,3,4}$ is a $3\!\times\!3$-dimensional submatrix of $\mathcal{H}$.
The \emph{Langevin equations for two spatial dimensions} are then given by
\begin{equation}
\begin{split}
\dot{\vec{r}}&=\vec{B}_{\mathrm{I}}+\beta\big(\mathrm{D}_{\mathrm{T}}(\vec{F}_{\mathrm{A}}-\Nabla_{\vec{r}}U+\vec{f})
-\vec{D}_{\mathrm{C}}(M-\partial_{\phi}U+\tau)\big)\;,\\
\dot{\phi}&=\beta\big(D_{\mathrm{R}}(M-\partial_{\phi}U+\tau)-
\vec{D}_{\mathrm{C}}\cdot(\vec{F}_{\mathrm{A}}-\Nabla_{\vec{r}}U+\vec{f})\big)
\end{split}\raisetag{12pt}
\label{eq:LangevinGLGII}
\end{equation}%
with the drift vector 
\begin{equation}
\vec{B}_{\mathrm{I}}(\phi)=B^{\parallel}_{\mathrm{I}}\hat{u}_{\parallel}+B^{\perp}_{\mathrm{I}}\hat{u}_{\perp}\;,
\end{equation}
which is in accordance with the interpretation of the solution of the Langevin equations as an It\={o} process \cite{KloedenP2006},
the translational short-time diffusion tensor
\begin{equation}
\begin{split}
\mathrm{D}_{\mathrm{T}}(\phi)&=D_{1}\hat{u}_{\parallel}\otimes\hat{u}_{\parallel}
+D_{2}(\hat{u}_{\parallel}\otimes\hat{u}_{\perp}+\hat{u}_{\perp}\otimes\hat{u}_{\parallel})\\
&\quad+D_{3}\hat{u}_{\perp}\otimes\hat{u}_{\perp}\;,
\end{split}
\end{equation}
and the coupling vector
\begin{equation}
\vec{D}_{\mathrm{C}}(\phi)=D^{\parallel}_{\mathrm{C}}\hat{u}_{\parallel}+D^{\perp}_{\mathrm{C}}\hat{u}_{\perp}\;.
\end{equation}
They involve only $8$ instead of $21$ shape-dependent parameters. 
These are the translational drift coefficients
{\allowdisplaybreaks\begin{align}%
B^{\parallel}_{\mathrm{I}}&=\frac{1}{\beta\eta}\Big(\big(\mathcal{H}^{-1}\big)_{24}-\big(\mathcal{H}^{-1}\big)_{15}\Big)\;,
\label{eq:Bp}\\
B^{\perp}_{\mathrm{I}}&=\frac{1}{\beta\eta}\Big(\big(\mathcal{H}^{-1}\big)_{16}-\big(\mathcal{H}^{-1}\big)_{34}\Big)\;,
\label{eq:Bs}%
\end{align}}%
the translational diffusion coefficients
{\allowdisplaybreaks\begin{align}%
D_{1}&=\frac{1}{\beta\eta}\big(\mathcal{H}^{-1}\big)_{33}=\frac{1}{\beta\eta}\big(\widetilde{\mathcal{H}}^{-1}\big)_{22}\;,\\
D_{2}&=\frac{1}{\beta\eta}\big(\mathcal{H}^{-1}\big)_{23}=\frac{1}{\beta\eta}\big(\widetilde{\mathcal{H}}^{-1}\big)_{12}\;,\\
D_{3}&=\frac{1}{\beta\eta}\big(\mathcal{H}^{-1}\big)_{22}=\frac{1}{\beta\eta}\big(\widetilde{\mathcal{H}}^{-1}\big)_{11}\;,
\end{align}}%
the coupling coefficients 
{\allowdisplaybreaks\begin{align}%
D^{\parallel}_{\mathrm{C}}&=\frac{1}{\beta\eta}\big(\mathcal{H}^{-1}\big)_{34}=\frac{1}{\beta\eta}\big(\widetilde{\mathcal{H}}^{-1}\big)_{23}\;,\\
D^{\perp}_{\mathrm{C}}&=\frac{1}{\beta\eta}\big(\mathcal{H}^{-1}\big)_{24}=\frac{1}{\beta\eta}\big(\widetilde{\mathcal{H}}^{-1}\big)_{13}\;,
\end{align}}%
and the rotational diffusion coefficient 
\begin{equation}
D_{\mathrm{R}}=\frac{1}{\beta\eta}\big(\mathcal{H}^{-1}\big)_{44}=\frac{1}{\beta\eta}\big(\widetilde{\mathcal{H}}^{-1}\big)_{33}\;.
\label{eq:DR}%
\end{equation}
Some of these eight coefficients are zero or equal, respectively, if the described Brownian particle is symmetric.
Table \ref{tab:Koeffizienten} gives an overview about possible symmetries of the particle's shape and the corresponding properties 
of the shape-dependent coefficients \eqref{eq:Bp}-\eqref{eq:DR}.

Although the Langevin equations \eqref{eq:LangevinGLGII} for two spatial dimensions are simpler than Eq.\ \eqref{eq:LangevinGLG},
they are still coupled nonlinear stochastic differential equations and thus not analytically solvable. 
However, if the external potential $U(\vec{r},\phi)$ is set to zero, the Langevin equations can be solved analytically
and the center-of-mass trajectory becomes a circle ($T=0$) or a logarithmic spiral ($T>0$) like in Ref.\ \cite{vanTeeffelenL2008}.
The analytical solution for $U(\vec{r},\phi)=f_{\parallel}=f_{\perp}=\tau=0$ is given by
\begin{widetext}
{\allowdisplaybreaks\begin{align}%
\begin{split}%
\vec{r}(t)&=\vec{r}_{0}
+\frac{\beta}{\omega}(F_{\parallel}D_{2}+F_{\perp}D_{3}-M D^{\perp}_{\mathrm{C}})
\big(\hat{u}_{\parallel}(\phi_{0}+\omega t)-\hat{u}_{\parallel}(\phi_{0})\big)\\
&\qquad\,-\frac{\beta}{\omega}(F_{\parallel}D_{1}+F_{\perp}D_{2}-M D^{\parallel}_{\mathrm{C}})
\big(\hat{u}_{\perp}(\phi_{0}+\omega t)-\hat{u}_{\perp}(\phi_{0})\big)\,,
\end{split}\\
\begin{split}%
\phi(t)&=\phi_{0}+\omega t
\end{split}%
\end{align}}%
\end{widetext}
with the angular velocity
\begin{equation}
\omega=\beta(M D_{\mathrm{R}}-D^{\parallel}_{\mathrm{C}}F_{\parallel}-D^{\perp}_{\mathrm{C}}F_{\perp})\;,
\end{equation}
the initial position $\vec{r}_{0}=\vec{r}(0)$, and the initial orientation $\phi_{0}=\phi(0)$.
If instead of $U(\vec{r},\phi)$ the coupling coefficients $D^{\parallel}_{\mathrm{C}}$ and $D^{\perp}_{\mathrm{C}}$ 
vanish, the Langevin equations become similar to the Langevin equations for the Brownian circle swimmer in Ref.\ \cite{vanTeeffelenL2008},
but with an additional non-diagonal coefficient $D_{2}$ in the parametrization of the translational diffusion tensor $\mathrm{D}_{\mathrm{T}}(\phi)$
and a more general driving force $\vec{F}_{\mathrm{A}}$, which is not necessarily parallel to the figure axis. 
For $D_{2}=D^{\parallel}_{\mathrm{C}}=D^{\perp}_{\mathrm{C}}=F_{\perp}=0$, the Langevin equations \eqref{eq:LangevinGLGII} are 
equivalent to the Langevin equation in Ref.\ \cite{vanTeeffelenL2008}.
An additional constraint on spherical particles that are only able to move along the $x_{1}$ axis leads to 
the Langevin equations for a spherical self-propelled particle on a substrate  \cite{tenHagenvTL2009}.
With similar simplifications it is also possible to derive the Langevin equations for spherical or anisotropic uniaxial 
self-propelled particles in two spatial dimensions that are discussed in Ref.\ \cite{tenHagenvTL2011}.

\subsection{\label{subsec:LangevinOrthotropic}Orthotropic particles}
Another possibility to simplify the Langevin equation \eqref{eq:LangevinGLG} considerably is the exclusive consideration of orthotropic particles.
All geometric bodies with three pairwise orthogonal planes of symmetry like spheres, spheroids (ellipsoids of revolution), biaxial (or triaxial) 
ellipsoids, cylinders, cuboids, and some prisms belong to this important class. In general, orthotropic particles have no 
translational-rotational coupling so that the coupling tensor $\mathrm{C}_{\mathrm{S}}$ vanishes. 
Furthermore, the translation tensor $\mathrm{K}$ and the rotation tensor $\Omega_{\mathrm{S}}$ are diagonal for orthotropic particles.
These properties of $\mathrm{K}$, $\mathrm{C}_{\mathrm{S}}$, and $\Omega_{\mathrm{S}}$ can be derived from the circumstance that the center of mass,
which should be chosen as reference point $\mathrm{S}$, and the mutual point of intersection of the three planes of symmetry of an orthotropic 
particle coincide \cite{HappelB1991}.
The vanishing of the coupling tensor $\mathrm{C}_{\mathrm{S}}$ results from the fact, that the point of intersection of the three pairwise 
perpendicular planes of symmetry of the particle is identical with the \emph{center of hydrodynamic reaction} for orthotropic bodies. 
With these considerations, the Langevin equation \eqref{eq:LangevinGLG} simplifies to the \emph{Langevin equations for orthotropic particles}:
\begin{equation}
\begin{split}
\dot{\vec{r}}&=\beta\:\!\mathrm{D}^{\mathrm{TT}}(\vec{\varpi})\big(\mathrm{R}^{-1}(\vec{\varpi})\vec{F}_{0}
-\Nabla_{\vec{r}}U+\mathrm{R}^{-1}(\vec{\varpi})\:\!\vec{k}_{1}\big)\;,\\
\vec{\omega}&=\beta\:\!\mathrm{D}^{\mathrm{RR}}(\vec{\varpi})\big(\mathrm{R}^{-1}(\vec{\varpi})\vec{T}_{0}
-\Nabla_{\vec{\varpi}}U+\mathrm{R}^{-1}(\vec{\varpi})\:\!\vec{k}_{2}\big)\\
&\quad+\Nabla_{\vec{\varpi}}\cdot\mathrm{D}^{\mathrm{RR}}(\vec{\varpi})\;.
\end{split}
\label{eq:LangevinGLGO}
\end{equation}
Here, the Gaussian white noises $\vec{k}_{1}$ and $\vec{k}_{2}$ are independent and defined as the first and second part of 
$\vec{k}=(\vec{k}_{1},\vec{k}_{2})$, respectively. 
The Langevin equations for spherical or uniaxial particles, that are considered in Ref.\ \cite{tenHagenvTL2011}, are special cases of the 
more general Langevin equations \eqref{eq:LangevinGLGO} for biaxial orthotropic particles. 
To be able to solve these Langevin equations analytically, it is at first necessary to neglect $\vec{k}_{1}$ and $\vec{k}_{2}$,
\ie, to consider the case $T=0$. 
A further negligence of the drive or the external potential leads to two special cases which are analytically solvable.

\subsubsection{Settling orthotropic passive particle:}
A particle without drive, \ie, with $\vec{F}_{0}=\vec{T}_{0}=\vec{0}$, moves only under the influence of the external potential 
$U(\vec{r},\vec{\varpi})$.
In the case of a constant gravitational field, only the constant external force $\vec{F}_{\mathrm{ext}}=-\Nabla_{\vec{r}}U$ acts on the particle 
and the external torque $-\Nabla_{\vec{\varpi}}U$ vanishes. 
The motion of such a sedimenting particle is well known from literature \cite{HappelB1991}.
It is characterized by a constant velocity $\dot{\vec{r}}$ and a constant orientation $\vec{\varpi}$: 
\begin{equation}
\begin{split}
\dot{\vec{r}}&=\beta\:\!\mathrm{D}^{\mathrm{TT}}(\vec{\varpi}_{0})\vec{F}_{\mathrm{ext}}=const.\;,\\
\vec{\varpi}&=\vec{\varpi}_{0}=const.
\end{split}
\end{equation}

\subsubsection{Self-propelled orthotropic particle:} 
\begin{figure}[ht]
\centering
\includegraphics[height=0.6\linewidth]{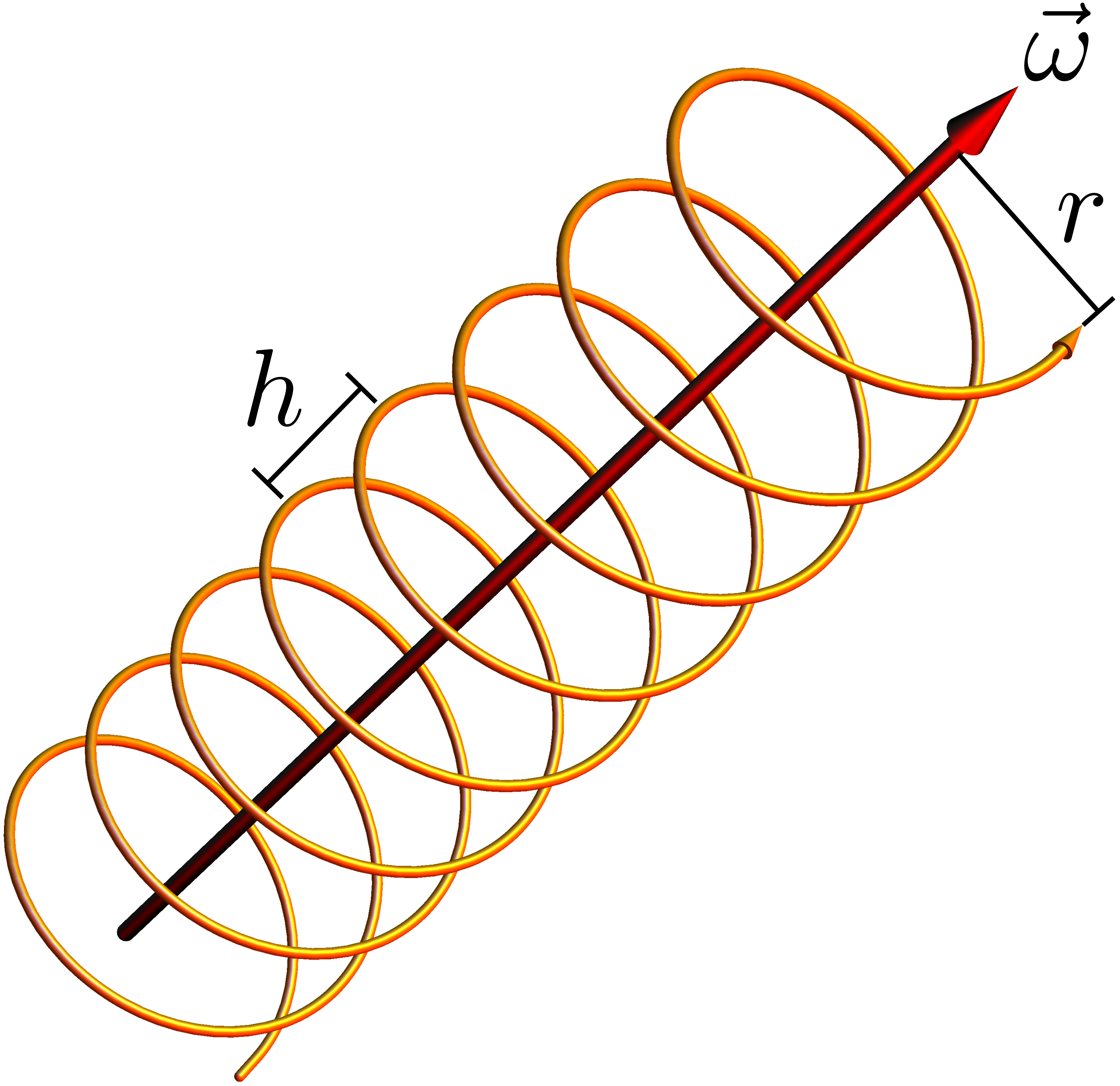}
\caption{The center-of-mass trajectory for $T=0$ and a constant external potential is a circular helix with radius $r$ and pitch $h$ that
evolves from the rotation of the orthotropic particle with the constant angular velocity $\vec{\omega}$.}
\label{fig:Helix}
\end{figure}

If the external potential $U(\vec{r},\vec{\varpi})$ is neglected instead of the drive in Eqs.\ \eqref{eq:LangevinGLGO} for $T=0$,
they describe the helical motion of an arbitrary orthotropic self-propelled particle in the absence of external and random forces and torques:
\begin{equation}
\begin{split}
\dot{\vec{r}}&=\mathrm{R}^{-1}(\vec{\varpi})\,\mathrm{K}^{-1}\frac{1}{\eta}\vec{F}_{0} \;,\\
\vec{\omega}&=\mathrm{R}^{-1}(\vec{\varpi})\,\Omega^{-1}_{\mathrm{S}}\frac{1}{\eta}\vec{T}_{0} \;.
\end{split}
\label{eq:LangevinNurAntrieb}
\end{equation}
These equations of motion are trivial in the body-fixed coordinate system, where
the velocities $\dot{\vec{r}}$ and $\vec{\omega}$ are constant.
Since the angular velocity $\vec{\omega}$ with the initial orientation $\vec{\varpi}_{0}=\vec{\varpi}(0)$ is constant in body-fixed coordinates, 
it is also constant in the space-fixed system.
This means that one expects a helix for the center-of-mass trajectory as it is known from the motion of 
protozoa like \emph{Euglena gracilis} \cite{JahnV1972} and bacteria like \emph{Thiovulum majus} \cite{SchulzJ2001}.
The analytical solution of Eq.\ \eqref{eq:LangevinNurAntrieb} is in fact the circular helix
\begin{equation}
\begin{split}
\vec{r}(t)&=\vec{r}_{0}+\frac{(\vec{\omega}\times\vec{v}_{0})\times\vec{\omega}}{\abs{\vec{\omega}}^{3}}\sin(\abs{\vec{\omega}}t) \\
&\quad+\frac{\vec{\omega}\times\vec{v}_{0}}{\abs{\vec{\omega}}^{2}}\big(1-\cos(\abs{\vec{\omega}}t)\big)
+\frac{\vec{\omega}\cdot\vec{v}_{0}}{\abs{\vec{\omega}}^{2}}\,\vec{\omega}\:\!t
\end{split}
\end{equation}
with axis 
\begin{equation}
\mathbb{A}=\vec{r}_{0}+\frac{\vec{\omega}\times\vec{v}_{0}}{\abs{\vec{\omega}}^{2}} + \vec{\omega}\:\!\R
\end{equation}
along $\vec{\omega}$, radius 
\begin{equation}
r=\frac{\abs{(\mathrm{K}^{-1}\vec{F}_{0})\times(\Omega^{-1}_{\mathrm{S}}\vec{T}_{0})}}{\eta^{2}\abs{\vec{\omega}}^{2}} \;,
\end{equation}
and pitch
\begin{equation}
h=2\pi\frac{\abs{(\mathrm{K}^{-1}\vec{F}_{0})\cdot(\Omega^{-1}_{\mathrm{S}}\vec{T}_{0})}}{\eta^{2}\abs{\vec{\omega}}^{2}} \;,
\end{equation}
where $\vec{r}_{0}=\vec{r}(0)$ is the initial position, 
$\vec{v}_{0}=\dot{\vec{r}}(0)=\mathrm{R}^{-1}(\vec{\varpi}_{0})\,\mathrm{K}^{-1}\vec{F}_{0}/\eta$ is the initial velocity,
and $\abs{\vec{\omega}}=\abs{\Omega^{-1}_{\mathrm{S}}\vec{T}_{0}/\eta}$ is the modulus of the angular velocity. 
This helical trajectory is shown schematically in Fig.\ \ref{fig:Helix}.
When also a constant gravitational field is taken into account, the helix becomes deformed and, for example, its cross section might become elliptic, 
but the axis of the helix remains a straight line. 
In the case $T>0$, where we have to consider the stochastic contributions in the Langevin equations \eqref{eq:LangevinGLGO}, 
it is, however, no longer possible to find analytical solutions. 
This case requires the usage of appropriate numerical integrators for stochastic differential equations and is studied in the next paragraph.

\section{\label{sec:NC}Numerical calculations}
\begin{figure}
\raggedright
\hskip0.25mm\includegraphics[width=0.95\linewidth]{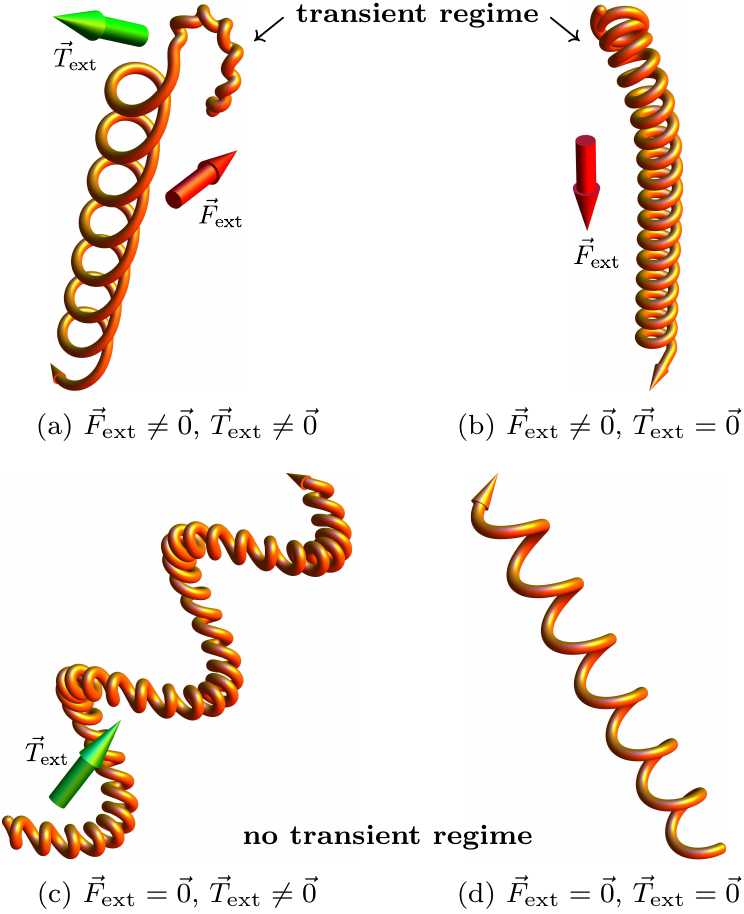}%
\caption{\label{fig:Klassifikation3D}Typical trajectories of arbitrarily shaped active particles in three spatial dimensions 
for constant vectors $\vec{F}_{0}\neq\vec{0}$, $\vec{T}_{0}\neq\vec{0}$ and temperature $T=0$.
The external force $\vec{F}_{\mathrm{ext}}=-\Nabla_{\vec{r}}U$ and torque $\vec{T}_{\mathrm{ext}}=-\Nabla_{\vec{\varpi}}U$ are constant, too. 
For a non-vanishing external force, the particle's center-of-mass trajectory starts with an irregular transient regime and 
changes into a periodic motion, as it is shown in plots (a) and (b). 
The general periodic motion (a), that is observed if there is also a non-vanishing external torque, reduces to a circular helix (b) parallel to 
the external force, if there is no external torque. The other two plots (c) and (d) show the situation for a vanishing external force, 
where a transient regime is not observed. 
There, the trajectory is either a \textit{superhelix}-like curve (c) parallel to a non-vanishing external torque or a circular helix (d), 
if there is no external torque.
However, the trajectories (a)-(c) can also be irregular. Straight trajectories, that are preceded by a transient regime if 
$\vec{F}_{\mathrm{ext}}\neq\vec{0}$, can be observed, too.}
\end{figure}
In general situations, where analytical solutions do not exist, the Langevin equation \eqref{eq:LangevinGLG} can only be investigated with 
the help of numerical methods. Appropriate numerical methods in increasing order of the truncation error are the Euler-Maruyama method, 
the Milstein method, and stochastic Runge-Kutta methods \cite{KloedenP2006}, which have to be derived in the It\={o} sense in order to be
compatible with the Langevin equations in this paper. 
If there are no thermal fluctuations ($T=0$), the stochastic Langevin equation \eqref{eq:LangevinGLG} becomes deterministic and a 
standard Runge-Kutta scheme of high order can be applied. 
The numerical results for $T=0$ and $T>0$, that are presented in the following, have been obtained by an explicit fourth-order deterministic 
Runge-Kutta scheme \cite{AbramowitzS1972,PressTVF1992,ArfkenW2005,Butcher2008} and by a multi-dimensional explicit stochastic 
Runge-Kutta scheme of weak order 2.0 for It\={o} stochastic differential equations \cite{KloedenP2006}, respectively. 
During the whole section, $\vec{F}_{0}$, $\vec{T}_{0}$, $\vec{F}_{\mathrm{ext}}=-\Nabla_{\vec{r}}U$, and $\vec{T}_{\mathrm{ext}}=-\Nabla_{\vec{\varpi}}U$
are assumed to be constant vectors, which do not depend on $\vec{r}$ or $\vec{\varpi}$.
Furthermore, arbitrary Brownian particles with a hydrodynamic translational-rotational coupling and orthotropic particles without a 
translational-rotational coupling in two and three spatial dimensions are considered for $T=0$ and $T>0$. 
In parallel to the previous section, this section is divided into a first subsection about the general Langevin equation for three spatial dimensions, 
a second subsection about two spatial dimensions, and a third subsection about orthotropic particles.

\subsection{Three spatial dimensions}
For arbitrarily shaped particles in three spatial dimensions, one finds various differently shaped trajectories as solutions of the Langevin equation 
\eqref{eq:LangevinGLG}. Figure \ref{fig:Klassifikation3D} gives a selection of typical trajectories that can be observed for arbitrarily shaped
particles with an arbitrary drive at $T=0$. 
In order to sample some typical solutions, we have randomly chosen the $21$ shape-dependent parameters for the matrix $\mathcal{H}$, 
the components of the internal force $\vec{F}_{0}$ and torque $\vec{T}_{0}$, the components of the external force $\vec{F}_{\mathrm{ext}}$ 
and torque $\vec{T}_{\mathrm{ext}}$, and the initial conditions. More than $100$ random parameter combinations were considered. 
In doing so, four different cases with vanishing and non-vanishing vectors for $\vec{F}_{\mathrm{ext}}$ and $\vec{T}_{\mathrm{ext}}$ were 
distinguished (see Fig.\ \ref{fig:Klassifikation3D}).
Depending on the choice of $\vec{F}_{\mathrm{ext}}$ and $\vec{T}_{\mathrm{ext}}$, the observed trajectories appeared to share common features 
and to be distinguishable into four different classes. The four trajectories that are shown in Fig.\ \ref{fig:Klassifikation3D} are representatives of
these classes. They can be characterized as follows:
If there are both an external force and an external torque, self-propelled particles that start with an 
irregular transient regime and end up in a simple periodic center-of-mass trajectory are usually observed [see Fig.\ \ref{fig:Klassifikation3D}(a)].
Note that the length of the initial transient regime as well as the periodicity length of the final periodic motion depend on the particular parameters
and may become rather long. 
In the case of no external torque, the periodic motion after the transient regime is a circular helix with its axis being parallel to the direction 
of the external force vector. This situation is illustrated in Fig.\ \ref{fig:Klassifikation3D}(b). 
The analog case of an external torque but no external force is schematically shown in Fig.\ \ref{fig:Klassifikation3D}(c).
There, a \textit{superhelix}-like curve with the orientation parallel to the direction of the external torque vector and without a preceding 
transient regime is observed. In contrast to the trajectory in Fig.\ \ref{fig:Klassifikation3D}(a), the complicated superhelix-like curve does not 
turn into a simpler periodic curve after some time, since there is no transient regime for $\vec{F}_{\mathrm{ext}}=\vec{0}$.
As fourth case, the motion in the absence of both external forces and torques is shown in Fig.\ \ref{fig:Klassifikation3D}(d).
It appears to be a circular helix. Also in this case, a transient regime is not observed. 
In the situation of Fig.\ \ref{fig:Klassifikation3D}(a) and (b), completely irregular trajectories appear, when the transient regime is very long.
Even in the transient-free situation of Fig.\ \ref{fig:Klassifikation3D}(c), irregular trajectories are observed, when the rotational frequency ratio 
between the immanent rotation of the self-propelled particle and the rotation due to the external torque is irrational. 
Furthermore, straight trajectories appear as a special case and are preceded by a transient regime if $\vec{F}_{\mathrm{ext}}\neq\vec{0}$.
A complete and detailed classification of all trajectories that can be observed in three spatial dimensions is, however, not possible,
since the number of the parameters that define the shape of the particle and all further relevant quantities like internal and external 
forces and torques is quite big, but this number is much smaller in two spatial dimensions, where a more detailed classification is possible.

\subsection{Two spatial dimensions}
\begin{table*}[ht]
\centering
\caption{\label{tab:Klassifikation2D}Detailed classification of the trajectories of arbitrarily shaped particles in two spatial dimensions for $T=0$
with respect to the symmetries that are summarized in Tab.\ \ref{tab:Koeffizienten}. 
For the external force, $\vec{F}_{\mathrm{ext}}\neq\vec{0}$ is chosen, since all trajectories become circles otherwise.
In the plots below, $\vec{F}_{\mathrm{ext}}$ is always oriented downwards, \ie, in the negative $x_{2}$-direction.
Internal and external torques are combined to the effective torque $M_{\mathrm{eff}}=M-\partial_{\phi}U$.}%
\includegraphics[width=\linewidth]{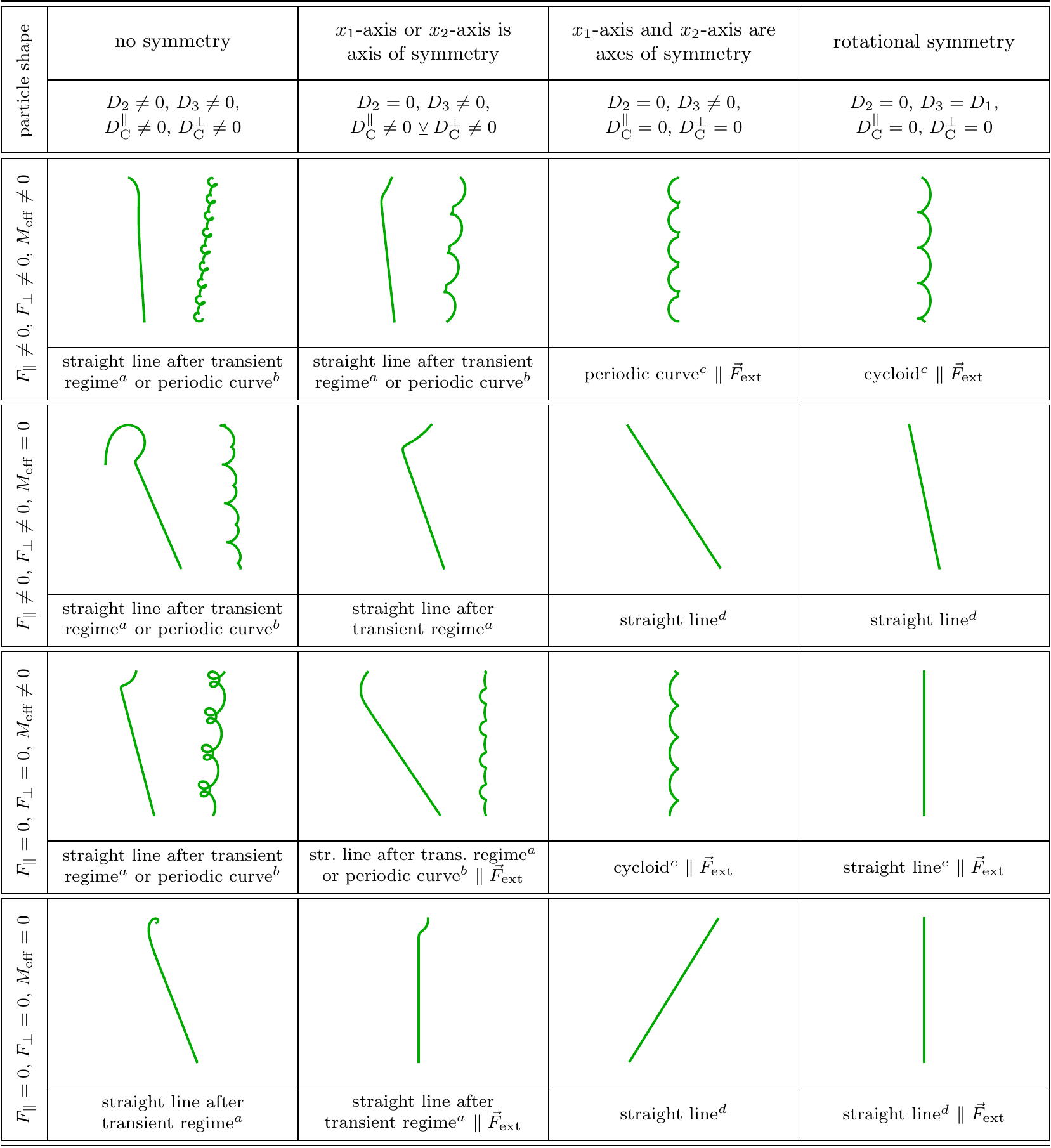}%
\begin{flushleft}
$^{a}$ The particle rotates monotonously until it reaches its final orientation. Then the angle $\phi$ remains constant.\\
$^{b}$ The angle $\phi$ and the center of mass of the particle describe periodic curves with the same periodicity.\\
$^{c}$ The particle rotates with a constant angular velocity, \ie, $\phi\propto t$.\\
$^{d}$ The orientation of the particle is constant: $\phi=const$.
\end{flushleft}
\end{table*}
The Langevin equations \eqref{eq:LangevinGLGII} for two spatial dimensions were solved analytically in Sec.\ \ref{subsec:LangevinIID}
for $U(\vec{r},\phi)=0$ and $T=0$. Here, we consider the case $T=0$, too, but now for a constant non-vanishing external force 
$\vec{F}_{\mathrm{ext}}=-\Nabla_{\vec{r}}U$, since the external force leads to various different non-trivial trajectories. 
The observed trajectories are classified with respect to the shape and the kind of self-propulsion of the particle in 
Tab.\ \ref{tab:Klassifikation2D}. 
Since the drift coefficients $B^{\parallel}_{\mathrm{I}}$ and $B^{\perp}_{\mathrm{I}}$ can be neglected for $T=0$, the
shape of the particle is only described with the six parameters $D_{1}$, $D_{2}$, $D_{3}$, $D^{\parallel}_{\mathrm{C}}$, 
$D^{\perp}_{\mathrm{C}}$, and $D_{\mathrm{R}}$, where $D_{1}$ and $D_{\mathrm{R}}$ can be set to one by a suitable rescaling of the 
length and time scales. For the remaining parameters, particles with a translational-rotational coupling and particles 
without a translational-rotational coupling as well as asymmetric particles, particles with one axis of symmetry, 
particles with two mutually perpendicular axes of symmetry, and isotropic particles with rotational symmetry are distinguished. 
Moreover, the constant parameters $F_{\parallel}$, $F_{\perp}$, and $M_{\mathrm{eff}}=M-\partial_{\phi}U$ are used to describe the 
self-propulsion of the particle. Four situations of a non-vanishing internal force $\vec{F}_{\mathrm{A}}$ and torque $M$, 
a drive only by either an internal force or an internal torque, and a passive particle with a vanishing drive are considered. 
A further distinction with respect to the external force and torque is not necessary, because the external force can always be chosen 
bigger than zero, since the case of a vanishing external force has been proven to lead to a trivial circular trajectory in  
Sec.\ \ref{subsec:LangevinIID}, and the external torque $-\partial_{\phi}U$ is already included in the effective torque $M_{\mathrm{eff}}$.
In general, straight lines with an aperiodic transient regime, arbitrary periodic curves, cycloids, and simple straight lines were found
as trajectories in two spatial dimensions by random choices of the parameters. 
These trajectories have still the basic features of their three-dimensional analogs, but are much simpler to describe.
It is only for particles with translational-rotational coupling, \ie, particles where at least one of the coefficients 
$D^{\parallel}_{\mathrm{C}}$ and $D^{\perp}_{\mathrm{C}}$ does not vanish, that the straight trajectories are preceded by an initial transient regime. 
These trajectories are characterized by a monotonous rotation of the particle when it starts moving and an ensuing rotation-free straight motion. 
In the transient regime, the particle rotates until the internal torque, the external torque, and the additional torque due to the 
translational-rotational coupling compensate each other. For active particles with translational-rotational coupling, also periodic 
trajectories are observed, where a canceling of the total torque does not happen during the initial rotation. 
This is not the case for symmetric particles with a vanishing effective torque $M_{\mathrm{eff}}$ and for passive particles, for which 
a periodic trajectory is not observed.
The motion of symmetric particles without translational-rotational coupling is always periodic or constant. 
In both cases, the trajectory is parallel to the direction of the external
force, if the effective torque $M_{\mathrm{eff}}$ is not zero. 
Without the effective torque, only straight trajectories are observed for these particles. 
Solely in the case of passive rotationally symmetric particles, the orientation of these straight trajectories is parallel to the external force.

\subsection{Orthotropic particles}
\begin{figure*}[ht]
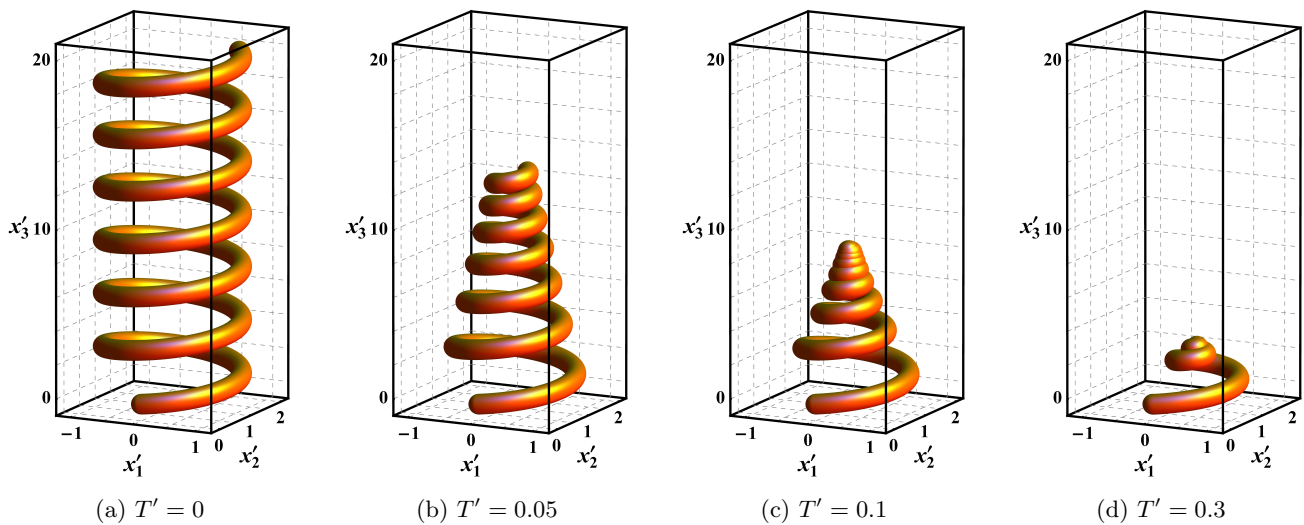

\centering
\begin{tabularx}{\linewidth}{YYYY}
\includegraphics[height=1.5\linewidth]{Abb3a}
&\includegraphics[height=1.5\linewidth]{Abb3b}
&\includegraphics[height=1.5\linewidth]{Abb3c}
&\includegraphics[height=1.5\linewidth]{Abb3d}\\
(a) $T'=0$&(b) $T'=0.05$&(c) $T'=0.1$&(d) $T'=0.3$\\
\end{tabularx}
\caption{Mean trajectories of a self-propelled orthotropic particle in the absence of external forces and torques for the dimensionless 
temperatures $T'=0$, $T'=0.05$, $T'=0.1$, and $T'=0.3$ (from left to right).
In the left two plots, the trajectories are shown for the time interval $0\leqslant t'\leqslant 40$, while the right two plots show mean 
trajectories with $0\leqslant t'\leqslant 80$.}
\label{fig:CS}%
\end{figure*}
\begin{figure}[ht]
\centering
\includegraphics[width=0.8\linewidth]{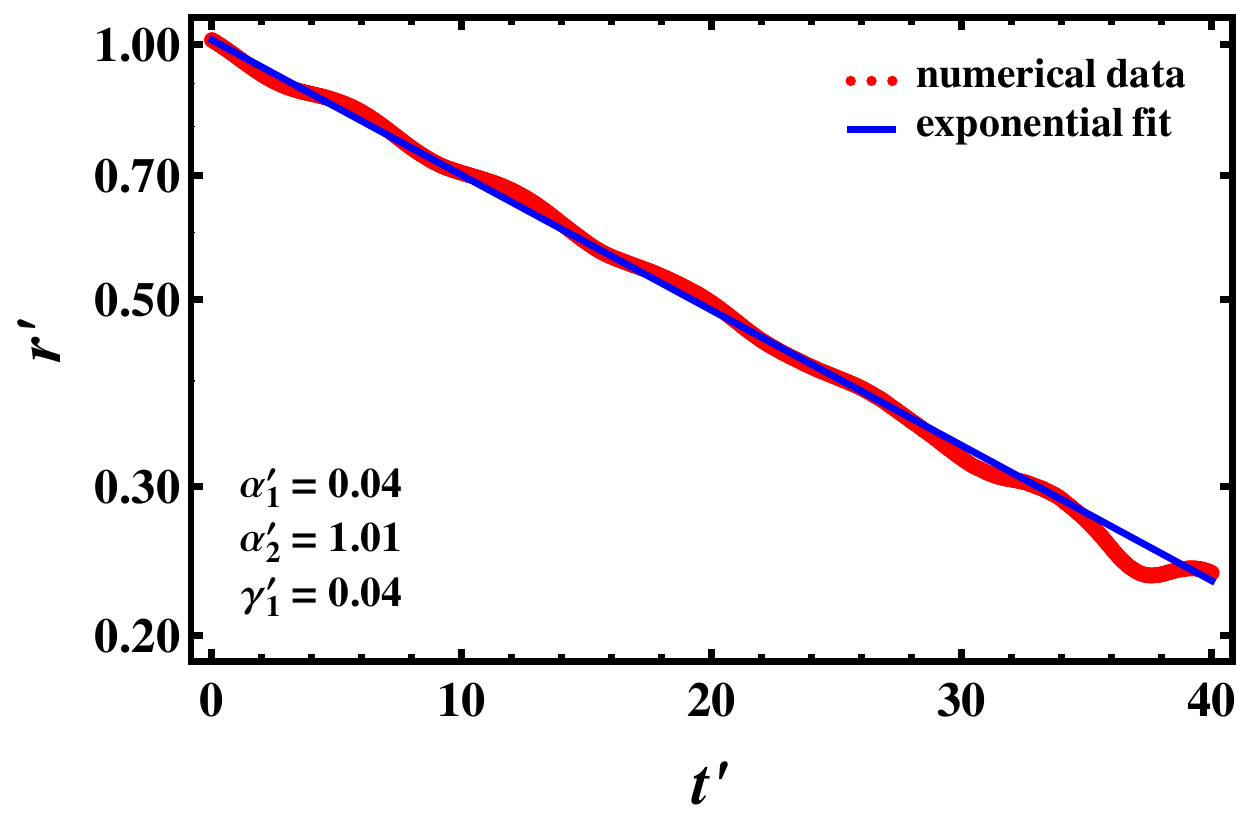}\\
(a) radius $r'(t')=r(t)/l_{\mathrm{c}}$ \\[3mm]
\includegraphics[width=0.8\linewidth]{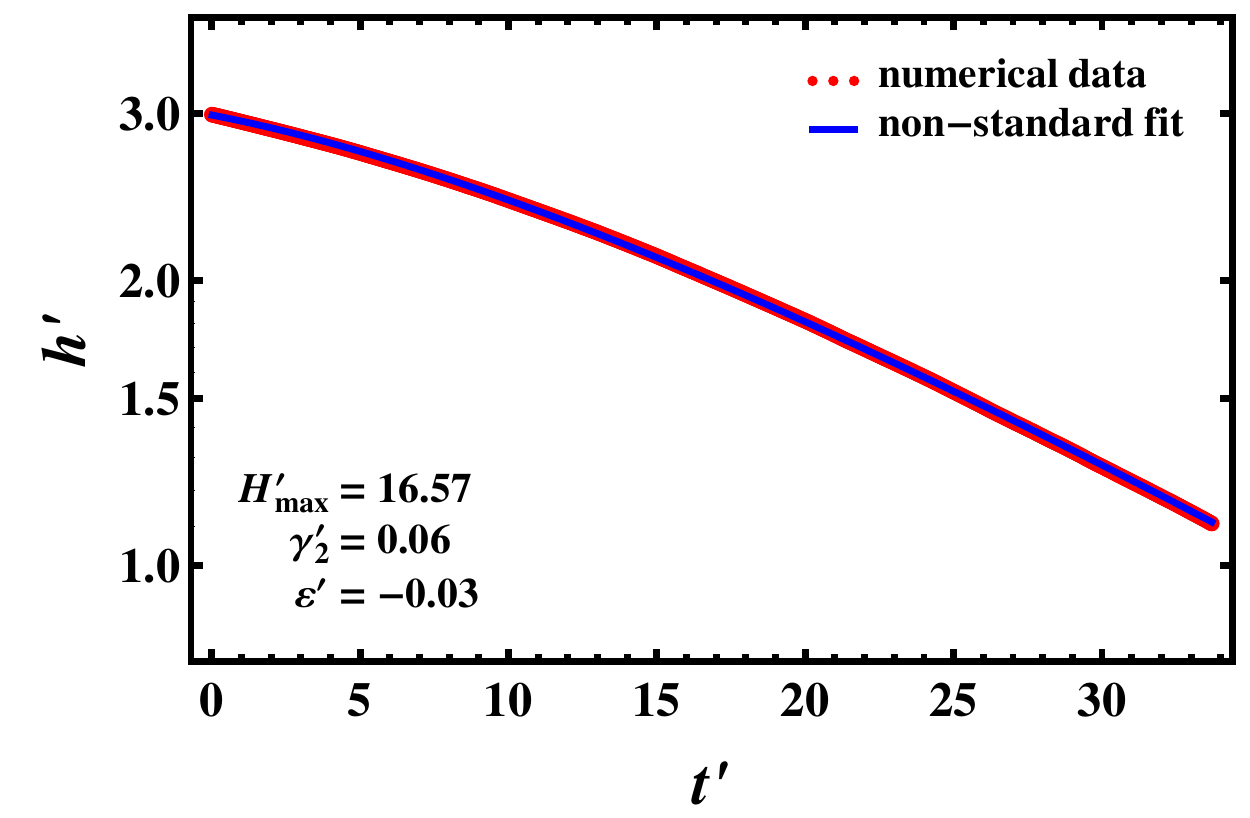}\\
(b) pitch $h'(t')=h(t)/l_{\mathrm{c}}$ \\[3mm]
\includegraphics[width=0.8\linewidth]{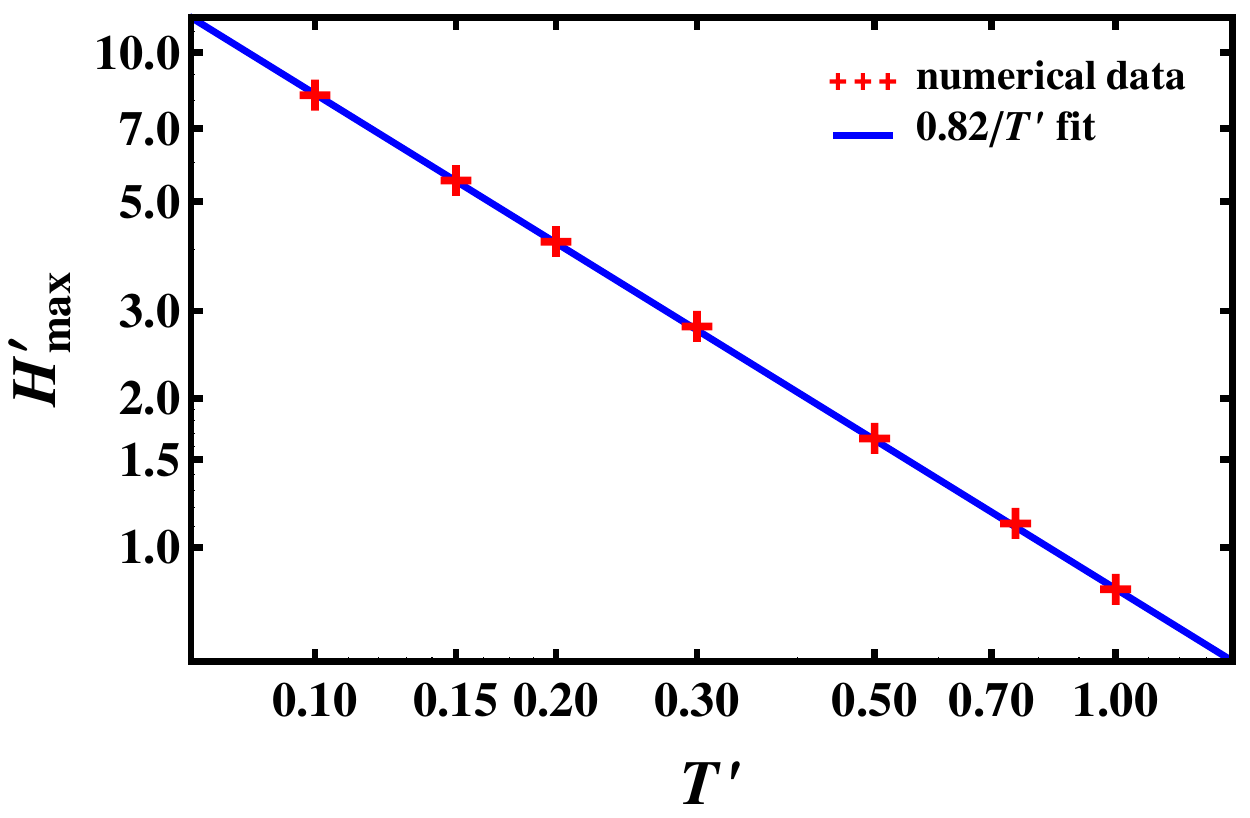}\\
(c) height $H_{\mathrm{max}}'(T')=H_{\mathrm{max}}(T)/l_{\mathrm{c}}$ 
\caption{(a) Radius and (b) pitch of a trajectory, that is shaped as a concho-spiral, decay 
exponentially with time $t'$. In the linear-logarithmic plots (a) and (b), the exponential decay is obvious for the radius, but not for the 
pitch, where the exponential function has a linear time-dependent prefactor [see Eq.\ \eqref{eq:hsa}].  
The numerical data (red dots following a curved line) for these plots were taken from the concho-spiral in Fig.\ \ref{fig:CS}(b). 
(c) The height of the concho-spirals is inversely proportional to the temperature $T'$.
The parameters for plot (c) are the same as in Fig.\ \ref{fig:CS}, but with more and different values for $T'$.
In each plot, a straight blue line corresponding to Eqs.\ \eqref{eq:rsaI}-\eqref{eq:Hmaxsa} was fitted to the numerical data.} 
\label{fig:rhHmax}%
\end{figure}
To regard the influence of thermal fluctuations on the motion of a biaxial self-propelled Brownian particle, the Langevin equations
\eqref{eq:LangevinGLGO} for orthotropic particles are solved numerically for $T>0$ in this section. 
For this purpose, first of all, characteristic quantities for the length scale, time scale, and force scale are chosen and the Langevin equations 
\eqref{eq:LangevinGLGO} are rescaled to dimensionless units. A suitable choice for the characteristic length $l_{\mathrm{c}}$,
the characteristic time $t_{\mathrm{c}}$, and the characteristic force $F_{\mathrm{c}}$ is 
{\allowdisplaybreaks
\begin{align}%
l_{\mathrm{c}}&=\sqrt{\frac{\lambda_{\mathrm{max}}(\mathrm{D}^{\mathrm{TT}})}{\lambda_{\mathrm{max}}(\mathrm{D}^{\mathrm{RR}})}}\;,\\
t_{\mathrm{c}}&=\frac{1}{\lambda_{\mathrm{max}}(\mathrm{D}^{\mathrm{RR}})}\;,\\
F_{\mathrm{c}}&=\frac{\eta\:\!l^{2}_{\mathrm{c}}}{t_{\mathrm{c}}}=\eta\,\lambda_{\mathrm{max}}(\mathrm{D}^{\mathrm{TT}})\;,
\end{align}}%
where $\lambda_{\mathrm{max}}(\,\cdot\,)$ denotes the biggest eigenvalue of the respective matrix.
These characteristic quantities are used to express the position $\vec{r}=\vec{r}'l_{\mathrm{c}}$, time $t=t't_{\mathrm{c}}$, 
forces $\vec{F}_{0}=\vec{F}_{0}'F_{\mathrm{c}}$, $\vec{F}_{\mathrm{ext}}=\vec{F}_{\mathrm{ext}}'F_{\mathrm{c}}$, 
torques $\vec{T}_{0}=\vec{T}_{0}'F_{\mathrm{c}}l_{\mathrm{c}}$, $\vec{T}_{\mathrm{ext}}=\vec{T}_{\mathrm{ext}}'F_{\mathrm{c}}l_{\mathrm{c}}$,
translation tensor $\mathrm{K}=\mathrm{K}'l_{\mathrm{c}}$, and rotation tensor 
$\Omega_{\mathrm{S}}=\Omega_{\mathrm{S}}'l^{3}_{\mathrm{c}}$ by the dimensionless position $\vec{r}'=(x_{1}',x_{2}',x_{3}')$, 
time $t'$, forces $\vec{F}_{0}'$, $\vec{F}_{\mathrm{ext}}'$, torques $\vec{T}_{0}'$, $\vec{T}_{\mathrm{ext}}'$, translation tensor $\mathrm{K}'$, and rotation tensor 
$\Omega_{\mathrm{S}}'$, respectively. 
In the rescaled Langevin equations, the parameter 
\begin{equation}
T'=\frac{2t_{\mathrm{c}}}{\eta\beta\:\!l^{3}_{\mathrm{c}}}=\frac{2}{\eta\beta}
\sqrt{\frac{\lambda_{\mathrm{max}}(\mathrm{D}^{\mathrm{RR}})}{\lambda^{3}_{\mathrm{max}}(\mathrm{D}^{\mathrm{TT}})}}\propto T
\end{equation}
appears as a dimensionless temperature. This parameter is varied and fluctuation-averaged trajectories are calculated for different temperatures
with fixed initial conditions $\vec{r}_{0}'$ and $\vec{\varpi}_{0}'$. 
The results for the case of vanishing external forces and torques, where the trajectory for $T'=T=0$ is known
from the analytical solution in Sec.\ \ref{subsec:LangevinOrthotropic} to be a circular helix, are shown in Fig.\ \ref{fig:CS}.
For this figure, we chose the dimensionless forces $\vec{F}_{0}'=(-0.5,0,3)$, $\vec{F}_{\mathrm{ext}}'=\vec{0}$, torques 
$\vec{T}_{0}'=(-1,0,0)$, $\vec{T}_{\mathrm{ext}}'=\vec{0}$, translation tensor $\mathrm{K}'=\Diag(1,2,3)$, rotation tensor 
$\Omega_{\mathrm{S}}'=\Diag(1,3,4)$, initial conditions $\vec{r}_{0}'=(x_{1,0}',x_{2,0}',x_{3,0}')=\vec{0}$, 
$\vec{\varpi}_{0}'=\vec{\varpi}_{0}=(0,\pi/2,0)$, and temperatures $T'=0,0.05,0.1,0.3$. 
It is apparent that the center-of-mass trajectory for $T'=0$ and the fluctuation-averaged center-of-mass trajectories for $T'>0$
have no transient regime in Fig.\ \ref{fig:CS}. This is also the case for non-vanishing constant external forces and torques and a 
general feature of orthotropic particles in contrast to less symmetric particles with a translational-rotational coupling.
In the presence of thermal fluctuations, the helical motion of the self-propelled orthotropic particle is damped exponentially with 
time and the fluctuation-averaged center-of-mass trajectory becomes a \emph{concho-spiral} \cite{Boyadzhiev1999}, whose radius and pitch
decay exponentially with time [see Fig.\ \ref{fig:rhHmax}(a) and (b)].
This result was confirmed by a fit of the numerical solutions with the general parametrization of a concho-spiral and agrees 
with the observation of a logarithmic spiral, also named \emph{spira mirabilis} by Jacob Bernoulli, in the two-dimensional special 
case of our Langevin equations, that is investigated in Ref.\ \cite{vanTeeffelenL2008}.
In the situation of Fig.\ \ref{fig:CS}, the axes of the concho-spirals are parallel to the $x_{3}$-axis and can be parametrized by 
\cite{Boyadzhiev1999,vanTeeffelenL2008}
{\allowdisplaybreaks
\begin{align}%
\begin{split}%
x_{1}'(t')&=x_{1,0}'+\alpha_{1}'\big(\cos(\phi_{0})-\cos(\phi(t'))e^{-\gamma_{1}'t'}\big)\\
&\quad-\alpha_{2}'\big(\sin(\phi_{0})-\sin(\phi(t'))e^{-\gamma_{1}'t'}\big) \;,
\end{split}\label{eq:PCSa}\\%
\begin{split}%
x_{2}'(t')&=x_{2,0}'+\alpha_{1}'\big(\sin(\phi_{0})-\sin(\phi(t'))e^{-\gamma_{1}'t'}\big)\\
&\quad+\alpha_{2}'\big(\cos(\phi_{0})-\cos(\phi(t'))e^{-\gamma_{1}'t'}\big) \;,
\end{split}\label{eq:PCSb}\\%
\begin{split}%
x_{3}'(t')&=x_{3,0}'+H_{\mathrm{max}}'\big(1-(1-\varepsilon't')e^{-\gamma_{2}'t'}\big) \;,
\end{split}\label{eq:PCSc}\\%
\begin{split}%
\phi(t')&=\phi_{0}+\omega't'
\end{split}\label{eq:PCSd}%
\end{align}}%
with the angular frequency $\omega'=\abs{\vec{\omega}'}=\abs{\Omega_{\mathrm{S}}'^{-1}\vec{T}_{0}'}$ 
[see Eqs.\ \eqref{eq:LangevinNurAntrieb}] and the dimensionless fit parameters 
$\alpha_{1}'$, $\alpha_{2}'$, $\gamma_{1}'$, $\gamma_{2}'$, $H_{\mathrm{max}}'$, and $\varepsilon'$.
Equations \eqref{eq:PCSa}, \eqref{eq:PCSb}, and \eqref{eq:PCSd} describe a logarithmic spiral, which is the trajectory of the 
two-dimensional circle swimmer in Ref.\ \cite{vanTeeffelenL2008}, while the parametrization \eqref{eq:PCSc} of the third spatial variable 
$x_{3}'(t')$ is here more general than in Ref.\ \cite{Boyadzhiev1999}.
In Eq.\ \eqref{eq:PCSc} there is an additional term $\propto\varepsilon'$, which makes sure that a helix is obtained as special case of the 
concho-spiral for $T'=0$, \ie, for $\gamma_{1}'=\gamma_{2}'=0$. This is, however, not the case for the parametrization in Ref.\ \cite{Boyadzhiev1999}.
Based on the parametrization \eqref{eq:PCSa}-\eqref{eq:PCSd}, radius and pitch of the concho-spirals can be derived. 
The fit parameters $\alpha_{1}'$, $\alpha_{2}'$, and $\gamma_{1}'$ determine the dimensionless radius $r'(t')=r(t)/l_{\mathrm{c}}$ of the concho-spirals:
{\allowdisplaybreaks
\begin{align}%
r'(t')&=r'_{0}\:\!e^{-\gamma_{1}'t'}\;,\label{eq:rsaI}\\
r'(0)&=r'_{0}=\sqrt{\alpha_{1}'^{2}+\alpha_{2}'^{2}}\;\,.
\label{eq:rsaII}%
\end{align}}%
Their dimensionless pitch $h'(t')=h(t)/l_{\mathrm{c}}$ depends on the remaining fit parameters $\gamma_{2}'$, $H_{\mathrm{max}}'$, 
and $\varepsilon'$:
\begin{equation}
\begin{split}
h'(t')&=H_{\mathrm{max}}'\:\!e^{-\gamma_{2}'t'}\!\Big((1-\varepsilon't')\big(1-e^{-\frac{2\pi}{\omega'}\gamma_{2}'}\big)\\
&\qquad\qquad\qquad\;\,+\varepsilon'\frac{2\pi}{\omega'}e^{-\frac{2\pi}{\omega'}\gamma_{2}'}\Big) \;.
\end{split}
\label{eq:hsa}%
\end{equation}
Numerical values for radius and pitch are shown in Fig.\ \ref{fig:rhHmax}(a) and (b). 
Furthermore, the height $H_{\mathrm{max}}(T)$ of the concho-spiral and its dimensionless analog 
$H_{\mathrm{max}}'(T')=H_{\mathrm{max}}(T)/l_{\mathrm{c}}$, 
defined as the distance from the initial position of the particle to its final position for $t'\to\infty$ measured along the axis 
of the concho-spiral, are finite and decrease monotonously, when the temperature is increased. 
For the numerical calculations that correspond to the results shown in Fig.\ \ref{fig:CS},
the inverse power law 
\begin{equation}
H_{\mathrm{max}}'(T')\approx 0.82 \,T'^{-1}
\label{eq:Hmaxsa}%
\end{equation}
was determined [see Fig.\ \ref{fig:rhHmax}(c)]. 
When there is additionally a constant external force, the helix for $T'=0$ as well as 
the concho-spirals for $T'>0$ are deformed and their cross-sections can become elliptic.

\section{\label{sec:conclusions}Conclusions and outlook}
In conclusion, we have studied the Langevin equation governing the motion of a driven Brownian spinning top
describing a self-propelled biaxial colloidal particle. The particle is driven both by internal 
and external forces and torques, which are constant in the body frame and in the lab frame, respectively. 
This equation is nontrivial due to the geometric biaxiality and the hydrodynamic translational-rotational coupling 
of the particle and can therefore only be solved numerically. In the special case of an orthotropic particle in the absence 
of external forces and torques, the noise-free trajectory is analytically found to be a circular helix and
the noise-averaged trajectory is a generalized \emph{concho-spiral}.
The noise-free trajectory is confirmed numerically to be more complex for translational-rotational coupling involving a 
transient irregular motion before ending up in simple periodic motion. By contrast, if the external force vanishes, no transient
is found and the particle moves on a \emph{superhelical} trajectory. We furthermore studied in detail the much simpler reduction
of the model to two spatial dimensions. In two spatial dimensions, the noise-free trajectories are classified completely and
circles, straight lines with and without transients, as well as cycloids and arbitrary periodic trajectories are found.

The Langevin equation derived here can be used as a starting point to describe more complex situations. 
As examples, we mention a particle in confinement of linear channels \cite{WensinkL2008,SchmidtvdGBWHF2008,ErbeZBKL2008} 
or cylindrical tubes \cite{vanTeeffelenZL2009,ChelakkotWG2010}. 
A sliding motion can be expected similar as in the two-dimensional reduction of our model
\cite{vanTeeffelenL2008}. In confinement, details of the propulsion mechanism are getting 
relevant since they result in different hydrodynamic interactions of the particle 
with the system boundaries \cite{DunkelPZY2010}.

A next level of complexity is given by an ensemble of swimmers, \ie, a finite number density,
which can interact either directly by excluded volume or via hydrodynamic interactions.
Recently, a dynamical density functional theory has been proposed for biaxial particles
\cite{WittkowskiL2011a} following the lines given for uniaxial particles \cite{RexWL2007}.
The collective properties of self-propelled biaxial particles are assumed to be rich 
including turbulent states, swarming, and jamming \cite{LeptosGGPG2009,PeruaniDB2006}

Finally, one can impose a non-vanishing solvent flow, like Couette shear flow 
\cite{tenHagenWL2011}, and study a self-propelled particle
there \cite{RafaiJP2010}. Our Langevin equation can be straightforwardly generalized to this situation \cite{MakinoD2004}
and may lead to new shear-induced tumbling phenomena of self-propelled biaxial particles.

\begin{acknowledgments}
We thank Gerhard N{\"a}gele, Ludger Harnau, Gerrit Jan Vroege, Holger Stark, Henricus H. Wensink, 
Michael Schmiedeberg, Borge ten Hagen, and Reinhard Vogel for helpful discussions.
This work has been supported by SFB TR6 (project D3).
\end{acknowledgments}

\bibliography{References}
\end{document}